# Van der Waals π Josephson junctions

Kaifei Kang[1*], Helmuth Berger[2], Kenji Watanabe[3], Takashi Taniguchi[3], László Forró[2,4], Jie Shan[1,5,6*] and Kin Fai Mak[1,5,6*]

[1] School of Applied and Engineering Physics, Cornell University, Ithaca, NY, USA
[2] Institute of Condensed Matter Physics, Ecole Polytechnique Fédérale de Lausanne, Lausanne, Switzerland
[3] National Institute for Materials Science, Tsukuba, Japan
[4] Stavropoulos Center for Complex Quantum Matter, Department of Physics, University of Notre Dame, Notre Dame, IN, USA
[5] Department of Physics, Cornell University, Ithaca, NY, USA
[6] Kavli Institute at Cornell for Nanoscale Science, Ithaca, NY, USA

*Email: kk726@cornell.edu; jie.shan@cornell.edu; kinfai.mak@cornell.edu

**Proximity-induced superconductivity in a ferromagnet can induce Cooper pairs with a finite center-of-mass momentum [1,2]. The resultant spatially modulated superconducting order parameter is able to stabilize Josephson junctions (JJs) with π phase difference in superconductor-ferromagnet heterostructures and realize "quiet" phase qubits [3-5]. The emergence of two-dimensional (2D) layered superconducting [6-9] and magnetic materials [10-12] promises a new platform for realizing π JJs with atomically sharp interfaces by van der Waals stacking [13]. Here we demonstrate a thickness-driven 0-π transition in JJs made of $NbSe_2$ (an Ising superconductor) with a $Cr_2Ge_2Te_6$ (a ferromagnetic semiconductor) weak link. By systematically varying the $Cr_2Ge_2Te_6$ thickness, we observe a vanishing supercurrent at a critical thickness around 8 nm, followed by a re-entrant supercurrent upon further increase in thickness. Near the critical thickness, we further observe unusual supercurrent interference patterns with vanishing critical current around zero in-plane magnetic field. They signify the formation of 0-π JJs (with both 0 and π regions) likely induced by the nanoscale magnetic domains in $Cr_2Ge_2Te_6$. Our work highlights the potential of van der Waals superconductor-ferromagnet heterostructures for the explorations of unconventional superconductivity and superconducting electronics.**

Superconductivity and ferromagnetism are two generally incompatible states of matter and do not coexist in a homogeneous material [1]. At a superconductor-ferromagnet interface, however, Cooper pairs can leak into the ferromagnet and give rise to exotic states, such as the Fulde–Ferrell–Larkin–Ovchinnikov (FFLO) state [1,2], spin triplet superconductivity [14,15], and topological superconductivity [16-18]. In particular, spin-singlet pairing in the ferromagnet with spin-split energy bands would force the Cooper pairs to acquire a center-of-mass momentum; a spatially periodic modulation in the superconducting order parameter known as the FFLO state is developed [1,2]. When the ferromagnet is sandwiched between two superconducting banks to form a JJ, the superconducting banks can have a 0 or π phase difference in the ground state depending on the thickness of the ferromagnetic weak link [1,2] (Fig. 1a). Specifically, when the superconducting order parameter at the two interfaces has opposite signs, a π JJ with negative Josephson coupling is stabilized. Although π JJs have been demonstrated in superconductor-ferromagnet-superconductor (SFS) heterostructures fabricated by evaporation and sputtering techniques [19,20], the recent discoveries of 2D layered superconducting [6-9] and magnetic materials [10-12] have provided a new platform to realize π JJs with atomically uniform thickness and sharp interfaces via van der Waals stacking [13].

In this study, we report π JJs and a 0-π transition in SFS heterostructures that are made of multilayer NbSe$_2$ (3-10 nm) and Cr$_2$Ge$_2$Te$_6$ (CGT for short, 3-13 nm) (Fig. 1b). Although such SFS heterostructures have been realized by recent experiments [21,22], clear evidence of FFLO physics and π JJs has not been demonstrated. Two-dimensional NbSe$_2$ is an Ising superconductor with a thickness dependent transition temperature around 3-7 K (Ref. [9]). The film thickness is substantially smaller than the out-of-plane penetration depth ($\approx$ 120 nm [23]) of bulk NbSe$_2$; an in-plane magnetic field thus penetrates the superconductor evenly without flux trapping. The extremely high in-plane upper critical field $\approx$ 30-40 T of 2D NbSe$_2$ (Ref. [9]) also renders the superconductor unperturbed by the weak magnetic fields examined in this study. Multilayer CGT is a hole-doped ferromagnetic semiconductor with a band gap around 0.38 eV and a Curie temperature $\approx$ 64 K (Ref. [24]). The in-plane saturation magnetic field is about 0.5 T (Extended Data Fig. 2), corresponding to relatively weak (out-of-plane) magnetic anisotropy among the known 2D magnetic materials [10]. Moreover, the remanent magnetization is negligible at low temperatures, which is consistent with the formation of nanoscale magnetic domains as revealed by a recent Lorentz TEM (transmission electron microscopy) study [22].

We examine the Josephson critical current of the junctions as a function of the in-plane magnetic field $B$ and the CGT thickness $d$. The critical current, $I_c$, is determined by measuring the differential resistance of the junction, $R = \frac{dV}{dI}$, versus a dc bias current $I$ ($V$ denoting the dc voltage across the junction). The junctions are nearly rectangular in shape. The magnetic field is applied along one of the principal axes. Because the junction dimension is negligible compared to the Josephson penetration depth (Methods), the junctions are point-like in space; the magnetic flux is uniform in the 2D plane and the Josephson supercurrent has negligible effects on the flux distribution. Unless otherwise specified, all results are obtained at $T = 1.6$ K, a temperature that is substantially lower than the superconducting transition temperature ($T_c \approx$ 6-7 K) of our devices. See Methods for details on the device fabrication and measurements.

The basic characterization of the JJs is illustrated in Fig. 1c-f with device JJ01 as an example (CGT thickness $d \approx$ 3.6 nm, total junction thickness 15 nm, and junction width $w \approx$ 2 μm). Figure 1d is the temperature dependence of the zero-bias differential resistance at $B = 0$ T. A sharp resistance drop to zero is observed at $T_c \approx 6.5$ K, below which a Josephson supercurrent is established. The $I - V$ characteristic and $R$ versus $I$ at 1.6 K are shown in Fig. 1e. Zero resistance ($R = 0$ and $V = 0$) is observed below the Josephson critical current $I_c \approx 62$ μA, at which sharp differential resistance peaks are also observed. The normal state resistance above $I_c$ is $R_n \approx 4$ Ω. The corresponding voltage jump at $I_c$ is $V_c = I_c R_n \approx 240$ μV, which is substantially smaller than the NbSe$_2$ superconducting gap $\Delta \approx 1$ meV (Ref. [25-28]) and consistent with proximity-induced superconductivity. No hysteresis is detected in the $I - V$ characteristic, showing that the JJ is overdamped [29]. These results suggest a diffusive junction [30] (see below for further discussions). A substantially larger voltage jump that shows hysteresis is observed at much larger bias current near 400 μA. The voltage jump (> 3 mV) exceeds the NbSe$_2$ superconducting gap. It corresponds to the critical current of the superconducting banks and is not related to the Josephson effect investigated in this study.

Figure 1f illustrates the supercurrent interference pattern under an in-plane magnetic field. The boundary separating the finite and zero resistance marks the magnetic-field dependence of the critical current $I_c$. The critical current is maximum near $B = 0$ T and shows a damped oscillatory dependence on field, which is well captured by a Fraunhofer pattern (dashed line). The oscillation period (70 mT) is consistent with $\Delta B = \frac{\Phi_0}{A_\perp}$, where $\Phi_0$ is the magnetic flux quantum and $A_\perp \approx 3 \times 10^{-14}$ m$^2$ is the cross-sectional area of the junction. The deviation from the ideal Fraunhofer pattern at larger magnetic fields is presumably caused by the internal magnetic field inside the ferromagnetic weak link. These results demonstrate the uniformity of the supercurrent and magnetic flux distributions in the 2D plane of the JJ. No magnetic hysteresis in the interference pattern can be detected for JJs with thin CGTs; this is consistent with negligible remanent magnetization in thin CGTs and negligible flux trapping in the JJ.

We now investigate the CGT thickness dependence of the JJ characteristics. We use the geometry-independent quantity $V_c = I_c R_n$ to characterize the Josephson coupling strength [20]. Figure 2 shows the temperature dependence of $V_c$ for zero-field cooled JJs of several representative CGT thicknesses. The superconducting transition temperature (marked by arrows) varies slightly among different devices because of the slight variations in the NbSe$_2$ thickness. In all cases, $V_c$ ($< \Delta$) becomes finite below $T_c$ and increases monotonically with decreasing temperatures. For JJs with thin CGT barriers (e.g. 3.6 nm and 5.9 nm), $V_c$ increases linearly near $T_c$ and saturates at low temperatures. The temperature dependence is well described by the Ambegaokar-Baratoff (AB) relation initially derived for tunnel JJs [30]. For thicker CGT (e.g. 9.9 nm), the temperature dependence substantially deviates from the AB relation; $V_c$ scales approximately as $V_c \propto \frac{T_c - T}{T}$ near $T_c$.

We approximate the saturated $V_c$ by its value at 1.6 K, the lowest measurement temperature in this study, and show its dependence on the CGT thickness $d$ (bottom axis) and layer number $N$ (top axis) in Fig. 3. As CGT thickness increases, $V_c$ first drops quickly, nearly reaches zero at $d_c \approx 8.4$ nm, followed by a revival, and vanishes again at $d \approx 12.3$ nm. (At $d \approx 8.4$ and 12.3 nm, $V_c$ drops below the measurement sensitivity that is limited by the electronic temperature in our experiment, see Methods). The observed oscillatory behavior of $V_c$ as a function of the barrier thickness in SFS JJs is a distinctive signature of a thickness-driven 0-$\pi$ transition in the current-phase relationship induced by FFLO physics [1,2]. In particular, JJs with $d \lesssim 8.2$ nm and $8.2 \lesssim d \lesssim 12.3$ nm have a 0 and $\pi$ phase difference between the two superconducting electrodes in the ground state, respectively. In contrast, oscillatory dependence of $V_c$ on temperature is not observed in any junctions (Fig. 2). The observed thickness dependence of $V_c$ in Fig. 3 can be well described by a damped oscillatory dependence derived for diffusive SFS junctions with $d$ substantially larger than the coherence lengths [31]

$$V_c = V_{c0} \cdot e^{-\frac{d}{\xi_{F1}}} \cdot \sin(\frac{d_c - d}{\xi_{F2}}). \qquad (1)$$

The best fit (solid line) corresponds to amplitude $V_{c0} \approx 0.94$ mV (which is comparable to the NbSe$_2$ superconducting gap $\Delta$) and coherence lengths $\xi_{F1} \approx 2.5 \pm 0.5$ nm and $\xi_{F2} \approx 2.4 \pm 0.5$ nm for CGT.

In general, the coherence lengths, $\xi_{F1,2} \approx \sqrt{\frac{\hbar v_f l}{E_x \mp \pi k_B T}}$, of a ferromagnetic weak link are determined by its scattering mean free path ($l$), out-of-plane Fermi velocity ($v_f$) and exchange interaction energy ($E_x$) with $\hbar$ and $k_B$ denoting the Planck and Boltzmann constants, respectively. Because the CGT Curie temperature 64 K is much higher than the NbSe$_2$ superconducting transition temperature $\approx$ 6 K, the coherence lengths are short below $T_c$, and the use of Eqn. (1) for diffusive SFS junctions is justified. The coherence lengths are also nearly temperature independent; for the entire range of $0 < T < T_c$ they change by about 0.1 nm, which is substantially smaller than the CGT interlayer distance of ~ 0.7 nm. Temperature-driven 0-π transitions are thus not expected at any fixed CGT thicknesses near $d_c$. The different temperature dependence of the JJ behavior with different weak link thicknesses (Fig. 2) can also be accounted for. For JJs with thin CGT (e.g. 3.6 nm), the junction length ($\approx d$) is comparable to the coherence lengths; the temperature dependence of $V_c$ for a diffusive JJ in this limit is shown to approximately follow the AB relation as in tunnel JJs [30]. As $d$ becomes substantially larger than the coherence lengths (e.g. 9.9 nm), $V_c$ is exponentially suppressed compared to $\Delta$ and can be shown to scale as $V_c \propto \frac{T_c - T}{T}$ near $T_c$ (Ref. [30]), as observed in experiment.

The discussion above treats the CGT weak link as a ferromagnetic conductor with a constant carrier density. The semiconducting nature of CGT and the charge transfer process at the NbSe$_2$-CGT interfaces are expected to give rise to a non-uniform charge density distribution inside the CGT weak link. Indeed, the junction resistivity ($\rho_n \equiv R_n A$) in the normal state scales with the CGT thickness $d$ exponentially (inset of Fig. 3, Methods), rather than linearly for a constant carrier density ($A$ is the overlap area of the two superconducting banks). The non-uniform charge density distribution also gives rise to a spatially varying Fermi velocity and coherence lengths in the CGT weak link, which can lead to a faster decay of $V_c$ with $d$ compared to the case of a spatially uniform case. Evidence of faster decay at the largest $d$'s is observed in Fig. 3 (the JJ with $d = 12.3$ nm has no measurable critical current).

Finally, we examine in Fig. 4 the supercurrent interference patterns in JJs with different CGT thicknesses. The differential resistance contour plots in Fig. 4a-d correspond to the forward field scanning direction. The extracted $I_c$ for both field-scanning directions are shown in Fig. 4e-h (arrows denoting the field scanning directions). For CGT thickness away from the critical thickness $d_c$ (i.e. 5.2 and 9.9 nm), a Fraunhofer-like pattern is observed with $I_c$ maximized near $B = 0$ T. A magnetic hysteresis due to the remanent magnetization and the associated flux lines in CGT is also observed. The hysteresis is more pronounced for the $d = 9.9$ nm device because it has a much larger cross-sectional area (corresponding to a smaller oscillation period). Remarkably, the interference pattern for CGT thicknesses just one monolayer away from $d_c$ (i.e. 7.7 and 9.1 nm) deviates substantially from the Fraunhofer pattern (Fig. 4b, c). The main central lobe is absent and a vanishing $I_c$ is observed near $B = 0$ T. For the reversed field scanning direction, a $I_c$ maximum near $B = 0$ T is restored (Fig. 4f, g). These interference patterns are robust for repeated forward and backward field scans within the illustrated magnetic-field range (Extended Data Fig. 3). They are, however, sensitive to magnetic-field cycles to higher fields and to thermal cycles to temperatures above the CGT Curie temperature (Extended Data Fig. 4 and 5). The sensitivity to magnetic-field and thermal cycles is only observed for JJs with $d \approx d_c$.

The vanishing $I_c$ for $d \approx d_c$ (for one scanning direction) indicates the presence of regions of opposite-direction supercurrent within the junction area that nearly cancels each other at $B = 0$ T. A finite magnetic field destroys the perfect supercurrent cancellation and yields two nearly symmetric critical current lobes around $B = 0$ T. Such unusual interference pattern is an experimental signature of the formation of a 0-π JJ, a special JJ that possesses spatially separated regions with a 0- and π-phase difference [32-34]. The magnetic hysteresis in the interference patterns and the sensitivity of the patterns to magnetic-field and thermal cycles suggest that the magnetic domain structure of the CGT weak link plays an important role in the formation of 0-π JJs. In particular, a different magnetic domain structure induced by field and thermal cycles destroys the supercurrent cancellation near $B = 0$ T and the 0-π JJ.

The possibility of stabilizing a 0-π JJ by magnetic domains in a ferromagnetic weak link of uniform thickness is suggested in Ref. [22,35]. The magnetic domain walls introduce additional spin-flip scatterings locally and therefore spatial modulations in the coherence lengths, which in turn modulate the sign of the Josephson coupling across the domain walls for a constant $d$ near $d_c$ and stabilize a 0-π JJ. This mechanism is distinct from what is demonstrated in SFS JJs with varying $d$ within the junction area [33,34]. It is plausible in our devices given the abundance of nanoscale magnetic domains in multilayer CGT [22]. Future studies are required to understand the microscopic origin of 0-π JJs formed for $d \approx d_c$.

In conclusion, we have demonstrated π JJs and a 0-π transition in NbSe$_2$/CGT/NbSe$_2$ SFS heterostructures by systematically varying the CGT thickness. Signatures of 0-π JJs are also observed for thicknesses near the critical thickness $d_c$. Because of the small band gap and unintentional hole doping in CGT, the JJs are in the diffusive limit. Future studies using large band gap ferromagnetic barriers, such as CrBr$_3$ and CrCl$_3$ (Ref. [10,28]), can potentially realize tunnel π JJs [36] and 'quiet' phase qubits [3-5].

**Methods**
**Device fabrications.** The NbSe$_2$/CGT/NbSe$_2$ heterostructures are assembled using the standard layer-by-layer transfer technique [13]. They are encapsulated in hexagonal boron nitride (hBN) flakes to prevent sample degradation under ambient conditions. Details of the fabrication process are reported in Ref. [13]. In short, multilayer NbSe$_2$ and CGT flakes are mechanically exfoliated onto Si substrates covered by a 300-nm thick oxide layer inside a glovebox filled with nitrogen gas; the hBN flakes are exfoliated in air. Their thicknesses are first estimated by optical reflection contrast, and characterized by atomic force microscopy (AFM). The selected flakes are picked up sequentially inside the glovebox by a polymer stamp made of a polycarbonate (PC) film on top of a dome-shaped polypropylene carbonate (PPC) droplet on a 0.5-mm-thick polydimethylsiloxane (PDMS) slab. The finished stack are released at 190 °C onto Si/SiO$_2$ substrates with pre-patterned platinum electrodes. Multiple devices with varying CGT thicknesses are examined in this study. A summary of all the devices is provided in Extended Data Figs. 1.

**Electrical measurements.** The differential resistance measurements are performed in an Oxford TeslatronPT closed cycle He-4 cryostat with a base temperature of ~ 1.6 K. The magnetic field is aligned parallel to the sample plane and along one of the principal axes of the nearly rectangular JJs. Flexible stainless steel coaxial cables are used as electrical connections to the devices. High frequency noise is filtered by three low-pass filters in series: a $\pi$-filter at room temperature, a homemade silver-epoxy filter and a low-temperature RC filter at about 10 K. They have cutoff frequencies at 30 MHz, 1 GHz and 30 kHz, respectively. These filters allow equilibration of the JJ electronic temperature with that of the exchange gas in a range of 1.6 - 7 K, limiting the measurable critical current to $I_c \sim 100$ nA. The JJs are biased by a DC current $I$, which is superposed by an AC current with RMS (root mean square) amplitude $dI = 50$ nA and frequency of 37 Hz. Both DC and AC voltage drops across the junction ($V$ and $dV$, respectively) are measured. The differential resistance is obtained as $R = \frac{dV}{dI}$.

**Estimate of the Josephson penetration depth.** Uniform supercurrent in a planar JJ is expected if the JJ width is much smaller than the Josephson penetration depth $\lambda_J = \sqrt{\frac{\Phi_0}{2\pi\mu_0 j_c d_{tot}}}$ (Ref. [37]). Here $\mu_0$ is the vacuum permeability and $j_c$ is the critical current density. Because the NbSe$_2$ thickness is substantially smaller than its out-of-plane penetration depth of ~ 120 nm, $\lambda_J$ is determined by the total thickness $d_{tot} \sim 10 - 30$ nm of the JJ. For JJs with thin CGT in our experiment ($d_{tot} \approx 15$ nm), the typical junction width is ~ 2 μm; $j_c$ is about $2 \times 10^7$ A/m$^2$. The Josephson penetration depth is $\lambda_J \sim 30$ μm, which is much larger than the junction width. For JJs with thicker CGT, the typical junction width increases to ~ 10 μm in order to measure the small $I_c$. Meanwhile, $j_c$ decreases exponentially with thickness, and correspondingly, $\lambda_J$ increases exponentially and remains much larger than the junction width. All JJs in this study are safely in the small junction limit with uniform supercurrent and magnetic flux through the entire junction area.

**Inhomogeneous carrier density in CGT.** Due to charge transfer at the NbSe$_2$/CGT interface, the interfacial CGT layers are expected to be more hole-doped than the interior layers when the CGT thickness is long compared to its screening length, which is inversely proportional to the square root of its thermodynamic density of states. (Because of the large density of states associated with the Cr d-orbitals and the large out-of-plane band mass due to the layered structure of CGT, a short screening length is expected.) The strong density dependence of the CGT resistivity near its mobility edge [24] therefore explains the observed exponential dependence of $\rho_n$ on $d$ in Fig. 3. (The substantial fluctuation in $\rho_n$ for devices with similar CGT thicknesses is likely caused by random variations in the unintentional doping density in the CGT flakes.)

# Figures

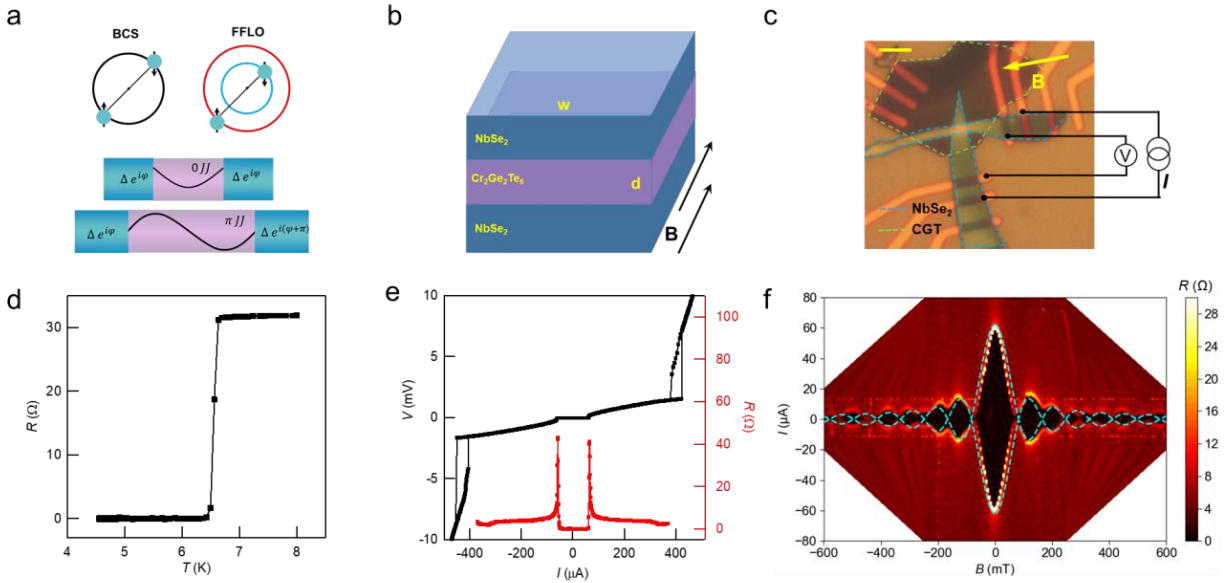

**Figure 1 | NbSe$_2$/Cr$_2$Ge$_2$Te$_6$/NbSe$_2$ Josephson junctions. a,** Top: schematic spin-singlet Cooper pairing in a Bardeen-Cooper-Schrieffer (BCS) superconductor (left) and in a ferromagnet with spin-split energy bands (right). A finite center-of-mass momentum is developed in the latter. The spin-up and -down bands are in red and blue, respectively. Bottom: Schematic 0 and π JJs. The former (latter) has superconducting order parameters of the same (opposite) sign at the NbSe$_2$/CGT interfaces. **b,** Schematic of a planar NbSe$_2$/CGT/NbSe$_2$ JJ. An in-plane magnetic field produces a magnetic flux through the cross-sectional area of the JJ. **c,** Optical micrograph of JJ01 (3.6 nm thick CGT) together with the measurement circuit. The arrow denotes the magnetic field direction. Scale bar is 5 μm. **d,** Temperature dependence of the zero-bias differential resistance at zero magnetic field for JJ01. Josephson effect is developed below ≈ 6.5 K. **e,** $I-V$ characteristic (black) and the bias dependent differential resistance (red) at 1.6 K and zero magnetic field. Sharp differential resistance peaks are observed at $I_c$. **f,** Dependence of the differential resistance on the bias current and magnetic field. Black region denotes zero resistance. The blue dashed line shows the expected Fraunhofer diffraction pattern.

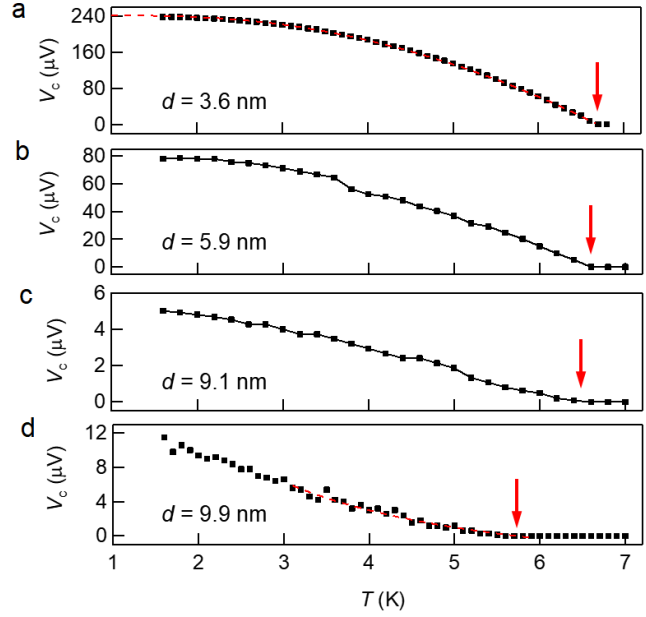

**Figure 2 | Temperature dependence of $V_c$. a-d,** Temperature dependence of the critical voltage $V_c = I_c R_n$ for JJs with selected CGT thicknesses of 3.6, 5.9, 9.1 and 9.9 nm. The expected temperature dependence described by the Ambegaokar-Baratoff relation and by $V_c \propto \frac{T_c - T}{T}$ is shown by the red dashed lines in **a** and **d**, respectively. Arrows mark the superconducting transition temperatures.

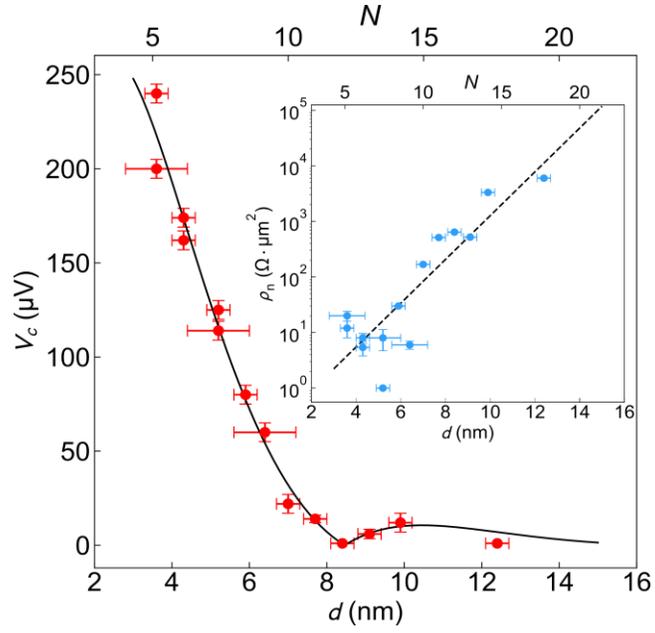

**Figure 3 | Thickness driven 0-π transition.** Dependence of the critical voltage $V_c$ on the CGT thickness $d$ (layer number $N$ in the top axis). Black line is the fit to the data according to Eqn. (1) in the main text. A 0-π transition is observed at the critical thickness ≈ 8.4 nm. The inset shows the thickness dependence of the normal state resistivity of the junctions. Near exponential dependence is observed. The thickness error bars are the standard deviations obtained from the AFM line profiles (Extended Data Fig. 1). The $V_c$ error bar is estimated as the standard deviation from different thermal cycles.

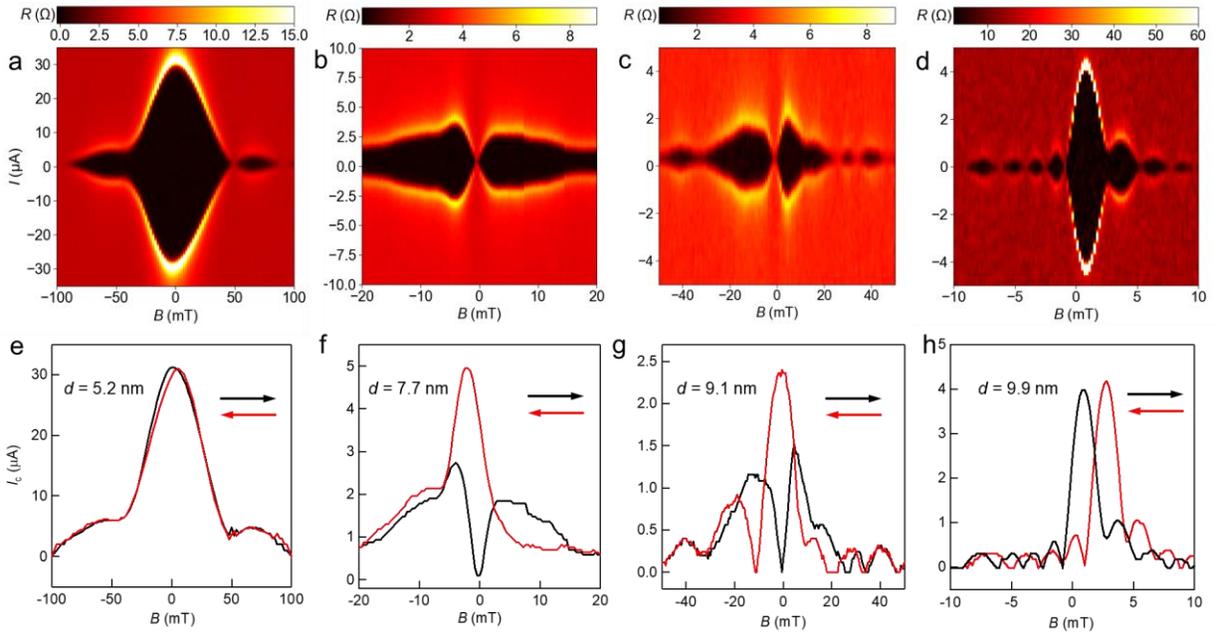

**Figure 4 | Thickness dependent supercurrent interference patterns. a-d,** Dependence of the differential resistance on the bias current and magnetic field for JJs with varying CGT thicknesses. Only the forward field scanning direction is shown. **e-h,** Extracted magnetic field dependence of the Josephson critical current for both field scanning directions (denoted by arrows). Whereas Fraunhofer-like patterns are observed for $d = 5.2$ and 9.9 nm, nearly vanishing critical current at zero magnetic field is observed for $d = 7.7$ and 9.1 nm for the forward field scanning direction.

**Extended data figures**

JJ 01
(*d* = 3.6 nm)

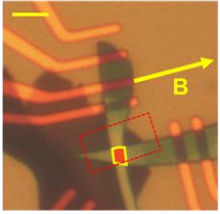 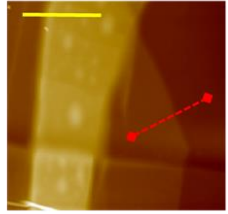 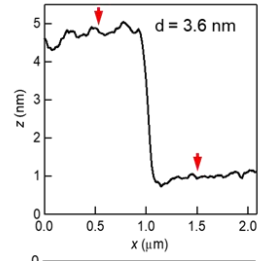 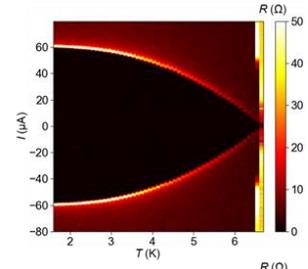

JJ 02
(*d* = 3.8 nm)

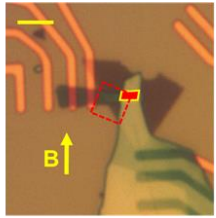 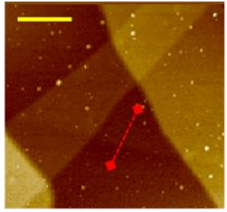 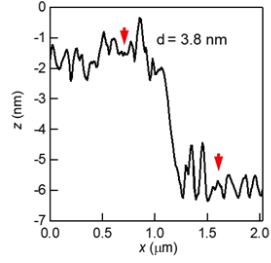 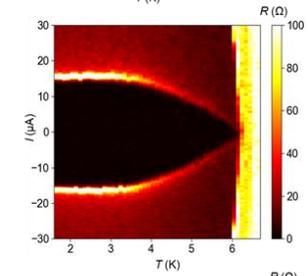

JJ 03
(*d* = 4.2 nm)

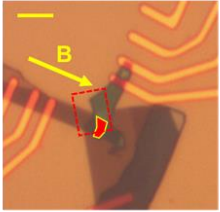 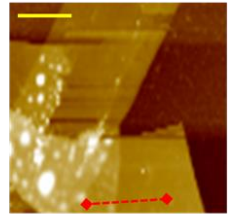 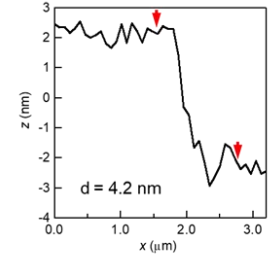 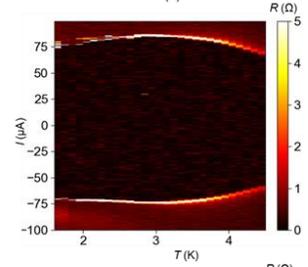

JJ 04
(*d* = 4.2 nm)

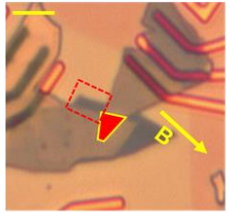 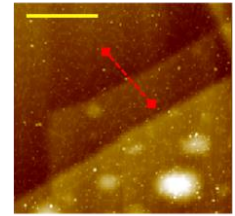 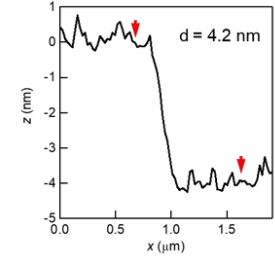 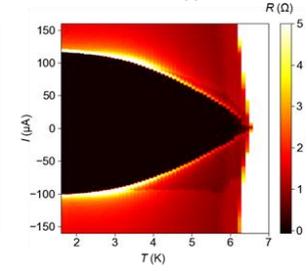

**JJ 05**
(*d* = 4.9 nm)

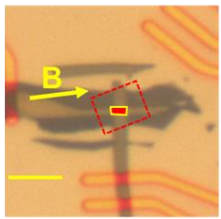 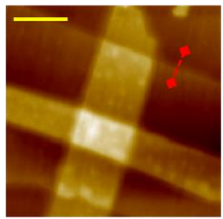 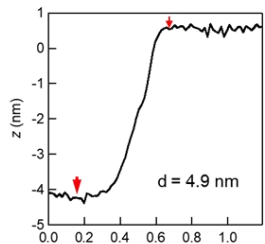 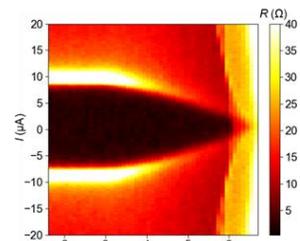

**JJ 06**
(*d* = 5.2 nm)

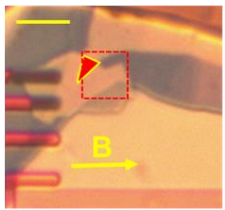 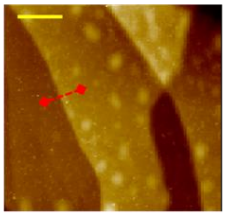 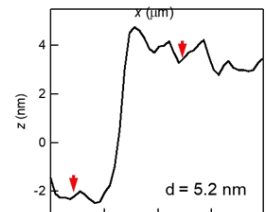 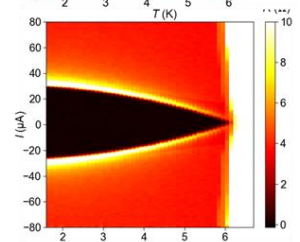

**JJ 07**
(*d* = 5.9 nm)

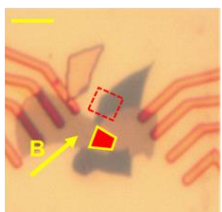 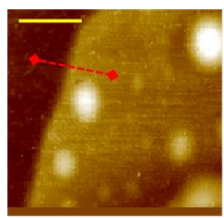 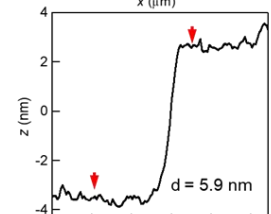 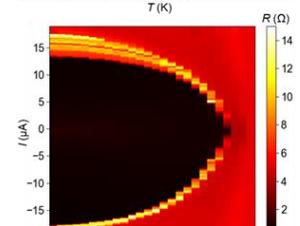

**JJ 08**
(*d* = 6.4 nm)

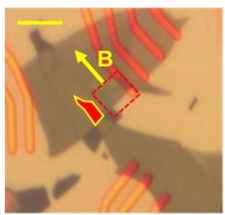 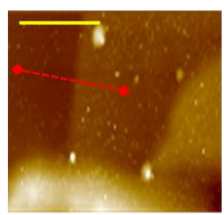 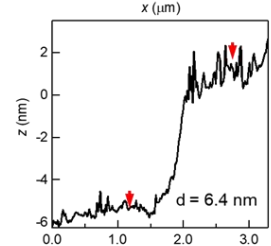 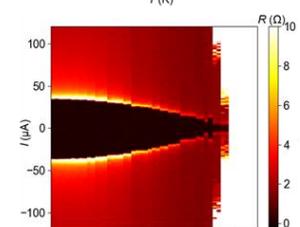

**JJ 09**
(*d* = 7.0 nm)

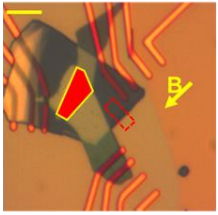 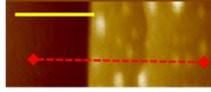 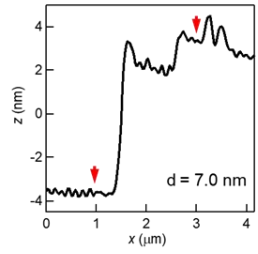 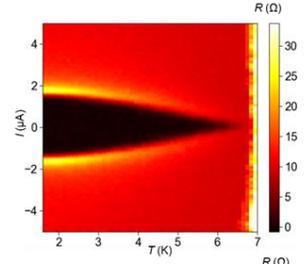

**JJ 10**
(*d* = 7.7 nm)

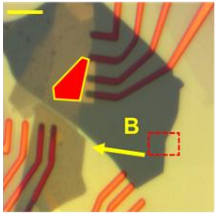 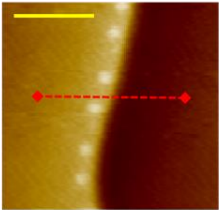 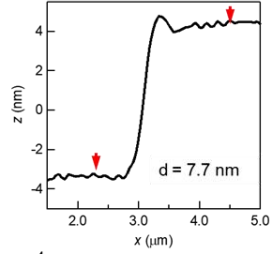 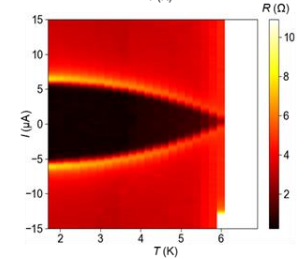

**JJ 11**
(*d* = 8.4 nm)

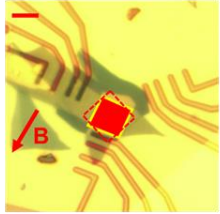 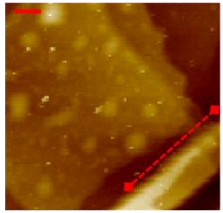 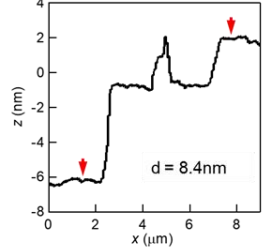 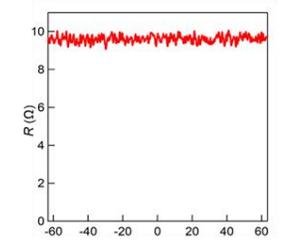

**JJ 12**
(*d* = 9.1 nm)

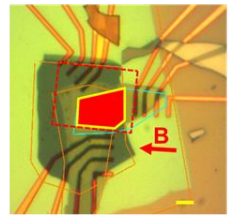 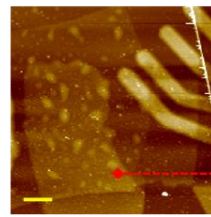 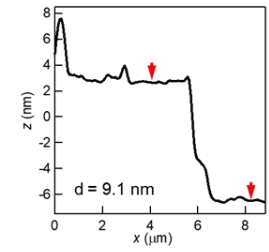 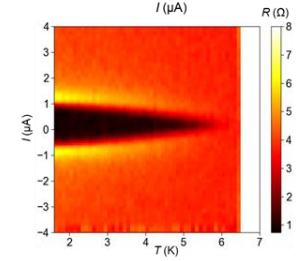

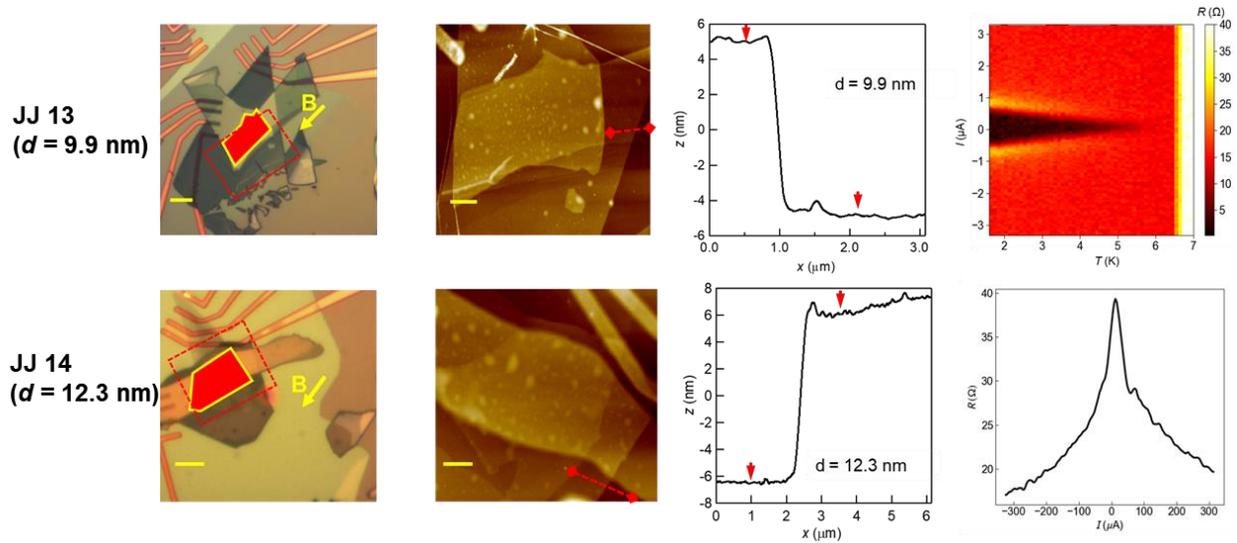

**Extended Data Figure 1 | Summary of all JJs (JJ01 – JJ14).** A summary of the device optical micrographs (first column), AFM images (second column), AFM line profiles (third column), and the dependence of the differential resistance on temperature and bias current under zero-magnetic-field cool down (forth column). Optical micrographs: The JJ area is marked red; the arrow denotes the magnetic field direction; scale bar is 5 μm; and the red-dashed box shows the area of the corresponding AFM image. AFM images: scale bar is 2 μm and the red dashed line denotes where the line profile is taken. AFM line profiles: the arrows denote the step height of the CGT weak link. Forth column: The boundary that separates zero and finite resistance determines the critical current. The critical current is below our measurement sensitivity ~ 100 nA for JJ11 and JJ14, in which the bias current dependence of the differential resistance is shown instead.

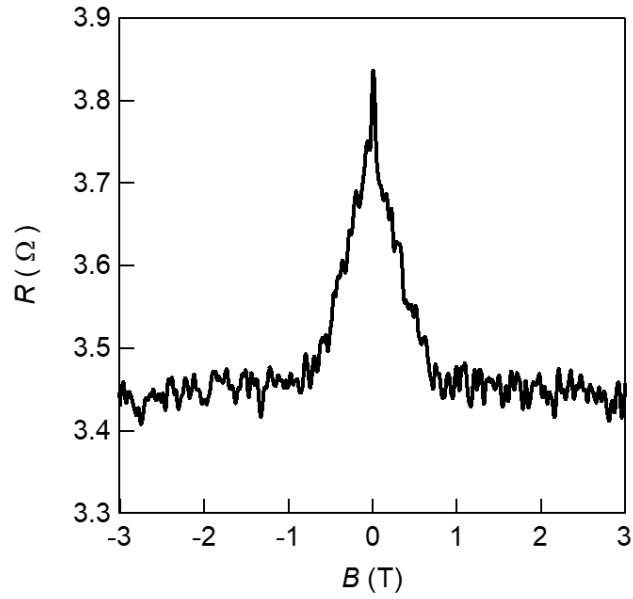

**Extended Data Figure 2 | Magneto-resistance at 8 K for JJ06 ($d$ = 5.2 nm).** Magnetic field dependence of the zero-bias differential resistance at 8 K, above the NbSe$_2$ superconducting transition temperature. A clear magneto-resistance below the in-plane saturation field (≈ 0.6 T) of CGT is observed.

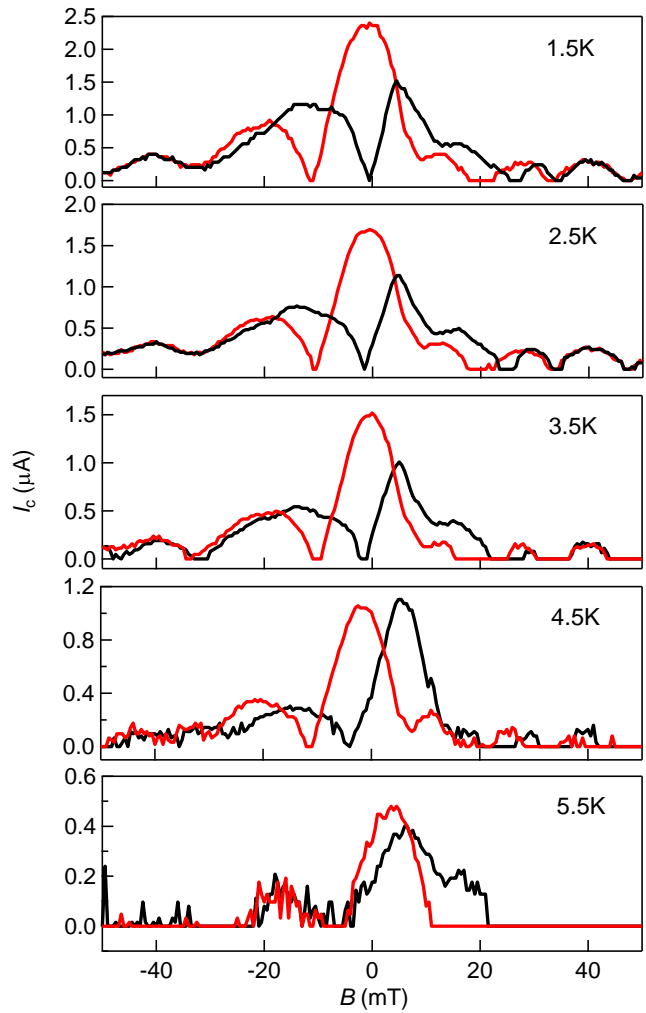

**Extended Data Figure 3 | Temperature dependent interference patterns for a 0-π junction (JJ12, $d$ = 9.1 nm).** Magnetic field dependence of the Josephson critical current at varying temperatures. The critical current decreases monotonically with increasing temperature. The characteristic interference pattern for 0-π JJs (see Fig. 4) is repeatable for multiple field scans and is observed up to about 3.5 K. A modified interference pattern starts to appear near 4.5 K, possibly caused by a change in the current-phase relation due to the modified spin-flip scattering rate near the domain walls at elevated temperatures.

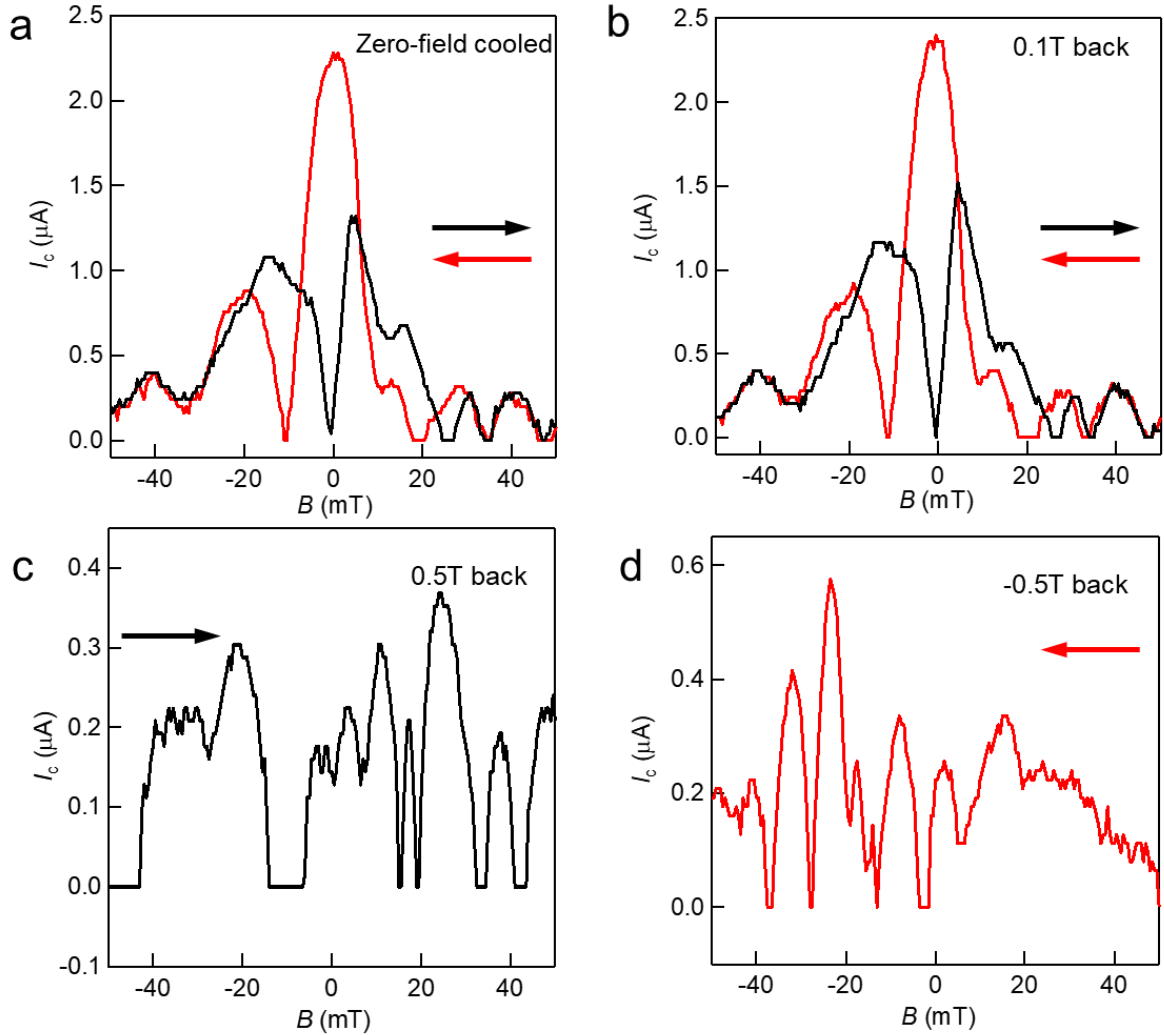

**Extended Data Figure 4 | Magnetic history dependence of the 0-π interference patterns (JJ12, *d* = 9.1 nm).** Magnetic field dependence of the Josephson critical current for both field scanning directions (marked by arrows) at 1.6 K obtained immediately after zero-field cool down (**a**) and after ramping the magnetic field to 0.1 T and back (**b**). The interference patterns are almost identical for the two cases, demonstrating the repeatability of the interference pattern and suggesting that the CGT magnetic domain structure remains almost identical after the 0.1 T field ramp. After ramping the magnetic field to ±0.5 T and back (**c,d**), however, the interference pattern becomes totally different. The magnetic domain structure is likely modified after a magnetic field ramp comparable to the in-plane saturation field (≈ 0.6 T). The result suggests the sensitivity of the supercurrent interference pattern on the magnetic domain structure.

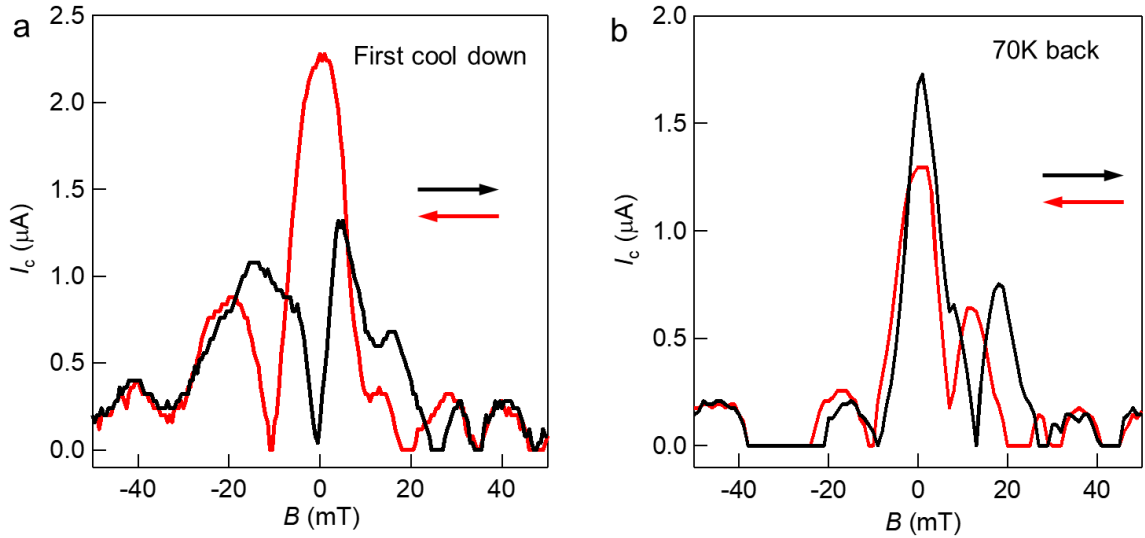

**Extended Data Figure 5 | Dependence of the 0-π interference patterns on thermal cycle (JJ12, $d$ = 9.1 nm).** Magnetic field dependence of the Josephson critical current for both field scanning directions (marked by arrows) at 1.6 K obtained after the first zero-field cool down (**a**) and after ramping the temperature to 70 K (above the Curie temperature) and cool back down under zero field (**b**). The interference patterns changed substantially. In particular, no 0-π JJ behavior is observed after the thermal cycle. The result suggests a substantial change in the CGT magnetic domain pattern after the thermal cycle, which modifies the interference pattern.